\documentclass[prb, aps, twocolumn, floatfix, superscriptaddress]{revtex4-2}
\usepackage{amsmath}
\usepackage{amssymb}
\usepackage{graphicx}
\usepackage{color}

\newcommand{\be}{\begin{equation}}
\newcommand{\ee}{\end{equation}}
\newcommand{\emb}{${\rm EuMg_2Bi_2}$}
\newcommand{\ems}{${\rm EuMg_2Sb_2}$}

\newcommand{\eg}{${\rm EuGa_4}$}
\newcommand{\ea}{${\rm EuAl_4}$}
\newcommand{\bea}{\begin{eqnarray}}
\newcommand{\eea}{\end{eqnarray}}
\newcommand{\bse}{\begin{subequations}}
\newcommand{\ese}{\end{subequations}}

\begin{document}

\title{Low-field magnetic anomalies in single crystals of the A-type square-lattice antiferromagnet EuGa$_4$}

\author{Santanu Pakhira}
\affiliation{Ames National Laboratory, Iowa State University, Ames, Iowa 50011, USA}
\author{D. C. Johnston}
\affiliation{Ames National Laboratory, Iowa State University, Ames, Iowa 50011, USA}
\affiliation{Department of Physics and Astronomy, Iowa State University, Ames, Iowa 50011, USA}

\date{\today}

\begin{abstract}

The body-centered-tetragonal antiferromagnet \eg\ was recently identified as a Weyl nodal-line semimetal that exhibits the topological Hall effect below its reported antiferromagnetic (AFM) ordering temperature $T_{\rm N}= 15$--16.5~K which we find to be $T_{\rm N} = 16.4(2)$~K\@. The Eu$^{+2}$ ions are located at the corners and body center of the  unit cell.  \eg\ exhibits A-type antiferromagnetic order below $T_{\rm N}$, where the Eu$^{2+}$ spin-7/2 moments are ferromagnetically aligned in the $ab$~plane with the Eu moments in adjacent Eu planes along the $c$~axis aligned antiferromagnetically.  Low-field magnetization versus field $M(H_{ab})$ data at $T=2$~K  with the field aligned in the $ab$~plane are reported that exhibit anomalous positive curvature up to a critical field at which a second-order transition occurs with $H_{c1}\approx 0.85$~kOe for ${\bf H}\parallel [1,1,0]$ and $\approx 4.8$~kOe for ${\bf H}\parallel [1,0,0]$. For larger fields, the linear behavior $M_{ab} = \chi(T_{\rm N})H_{ab}$ is followed until the critical field $H^{\rm c}_{ab}$ is reached at which all moments become aligned with the applied field.   A theory is formulated for $T=0$~K  that fits the observed $M(H_{ab})$ behavior at $T=2$~K well, where domains of A-type AFM order with fourfold rotational symmetry occur in the AFM state in zero field. The moments in the four domains reorient to become almost perpendicular to ${\bf H}_{ab}$ at $H_{c1}$, followed by increasing canting of all moments toward the field with increasing field up to the critical field $H^{\rm c}_{ab}$ which is reported to be 71 kOe, at which all moments become  aligned parallel to the field.  A first-order transition in $M(H_{ab})$ at $H_{ab}=H_{\rm c1}$ is predicted by the theory for $T=0$~K when ${\bf H}_{ab}$ is at a small angle from the [1,0,0] or [1,1,0] symmetry-axis directions.

\end{abstract}

\maketitle

\section{Introduction}

Antiferromagnets are fundamentally interesting owing to their various spin arrangements as well as their technological applications in spintronics, spin valves, magnetological devices, and spin-wave-based information technologies~\cite{Jungwirth2018, Park2011, Fert2017, Tokura2021}.  Recently, many antiferromagnetic (AFM) compounds have also been discovered to host nontrivial topological electronic and spin states~\cite{Otrokov2019, Rahn2018, Hirschberger2016, Kurumaji2019}.  Understanding the magnetic interactions in these materials are important for their further development and discovery of new materials.  The magnetic ordering in those materials are primarily determined by the interplay of exchange interaction, magnetic anisotropy energy, and any kind of disorder present in the system.  In particular, magnetocrystalline and magnetic-dipole anisotropies play a crucial role in tuning the spin arrangements in different AFM materials.

\begin{figure}
\includegraphics [width=1.75in]{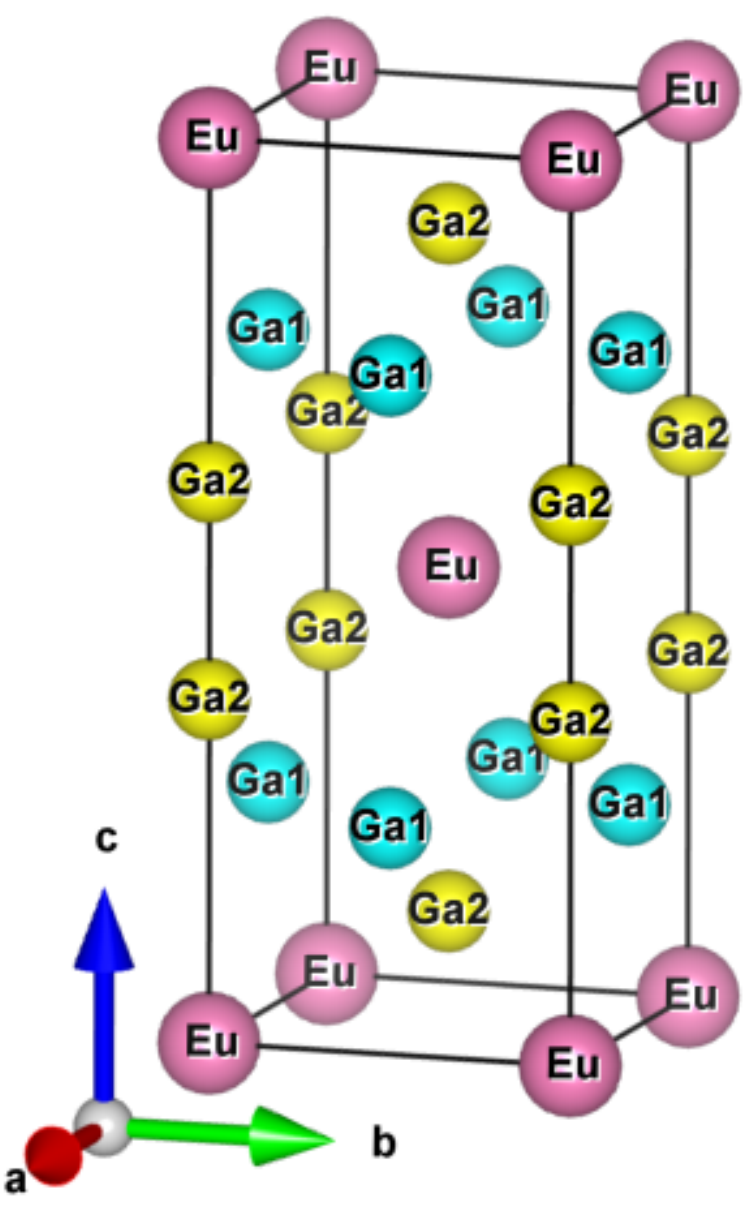}
\caption {Body-centered-tetragonal crystal structure of \eg. The Eu atoms form square lattices in the $ab$~plane.}
\label{Fig_structure}
\end{figure}

Among these materials, Eu-based antiferromagnets have been of significant interest recently due to the complex interplay of magnetism and topological states~\cite{Rahn2018, Jo2020, Riberolles2020, Li2019}.  Eu$X_4$-type of materials ($X$ = Al, Ga) constitute one such family which is generating significant interest due to the recent observation of the topological Hall effect (THE) and related phenomena in these materials~\cite{Shang2021, Zhang2022, Moya2021}.  They crystallize in the body-centered-tetragonal (bct) BaAl$_4$-type crystal structure (Fig.~\ref{Fig_structure}) with space group $I4/mmm$~\cite{Kneidinger2014}, where the Eu atoms in each $ab$-plane layer form a square lattice and are known to exhibit a rich variety of magnetic and electronic properties. For example, EuAl$_4$ orders antiferromagnetically below $T_{\rm N} = 15$~K along with a CDW transition at \mbox{ $T_{\rm CDW} = 140$~K\@~\cite{Araki2014, Nakamura2014, Nakamura2015, Stavinoha2018, Ramakrishnan2022}.}  The CDW transition is suppressed to $T = 0$~K by the application of a pressure of 2.5~GPa~\cite{Nakamura2015}.  The isovalent analogue EuGa$_4$ also orders antiferromagnetically below $T_{\rm N} \approx 16$~K and a CDW is only observed at $T_{\rm CDW} = 105$~K under the application of a pressure $p = 0.75$~GPa~\cite{Nakamura2015, Nakamura2013}.  A THE is also evidenced in EuAl$_4$ coexisting with CDW order~\cite{Shang2021}.  Although magnetic spin reorientation and multiple metamagnetic transitions were observed earlier in EuGa$_2$Al$_2$~\cite{Stavinoha2018}, a recent observation of the THE and incommensurate magnetic order suggest the presence of a field-induced skyrmion-like topological spin texture in this compound~\cite{Moya2021}.  A lack of inversion symmetry in non-centrosymmetric materials with Dzyaloshinskii-Moriya (DM) interactions was initially thought to be the key ingredient for stabilizing this spin texture.  However, observations of a skyrmionic phase in centrosymmetric materials~\cite{Kurumaji2019, Khanh2020, Hirschberger2019} have challenged the understanding and mechanism of this spin-texture formation.  Contemporary theoretical modeling suggests that the interplay of different spin interactions and anisotropy may play a crucial role in the formation of a topological spin texture in centrosymmetric materials~\cite{Hayami2021}.  Thus, to understand the mechanism of complex spin texture and its field-induced evolution, it is necessary to study the magnetic properties and anomalous behavior in antiferromagnets with small anisotropy.

EuGa$_4$ exhibits giant magnetoresistance (MR) and THE with a possibility of magnetic skyrmions~\cite{Zhang2022, Zhu2022}.  Recently, the observation of large transverse MR in this semimetal is explained due to the presence of Weyl nodal-line (NL) states and magnetic-field-induced Landau quantization~\cite{Lei2022}.  As reported earlier, EuGa$_4$ exhibits collinear A-type antiferromagnetic (AFM) order below $T_{\rm N} \approx 16$~K, where the Eu atoms are ferromagnetically aligned along the $ab$-planes and adjacent FM planes along the $c$~axis are aligned antiferromagnetically~\cite{Nakamura2014, Nakamura2015, Nakamura2013, Zhu2022, Kawasaki2016, Yogi2013}.  Although a noncollinear magnetic structure is favorable for skyrmion-like texture formation, the possibility of this texture in collinear EuGa$_4$ is quite intriguing where anisotropy can play an important role.  

Previous magnetic studies on \eg\ mostly focused on the magnetic ground state and the high-magnetic-field behavior, while the low-field behavior and the effect of anisotropy was hardly explored.  However, magnetization $M$ versus applied magnetic field $H$ isotherm measurements of a crystal with the field along the [1,0,0] direction at $T=2$~K revealed positive curvature up to a field $H_{\rm d} = 5$~kOe, above which $M(H)$ was linear up to the critical field $H^{\rm c}_{[1,0,0]} = 71$~kOe~at which all moments become parallel to the field, whereas for the $c$-axis field, $M(H)$ was linear over the whole field range where $H^{\rm c}_{[0,0,1]}=72$~kOe (nearly isotropic)~\cite{Nakamura2013}.  The authors suggested that this behavior was somehow associated with AFM domains that evolved into a single domain at $H_{\rm d}$, and found that $H_{\rm d}$ decreased smoothly to zero on heating to $T_{\rm N}$.

Here, we report studies of the magnetic-field evolution of the AFM ground-state spin texture at $T=2$~K in detail emphasizing the low-field region.  We found that although the $c$-axis magnetization increases linearly with the applied field $H$, as expected and found for a A-type AFM, a nonlinear $M(H)$ response at low fields was observed for the [1,0,0] field direction as previously found  in Ref.~\cite{Nakamura2013}.  Interestingly, we found that the low-field $(ab)$-plane nonlinearity differs significantly for fields in the [1,0,0] and [1,1,0] directions.  On the basis of our temperature- and magnetic-field-dependent magnetic measurements complemented with theoretical analyses, we conclude that the ground-state A-type AFM structure consists of four AFM domains having fourfold rotational symmetry associated with the fourfold ${ab}$-plane magnetic anisotropy of the ferromagnetic $ab$-plane layers.  We propose a theory in which, with increasing field in the $ab$~plane, the moments in each domain initially cant to become nearly perpendicular to the field at a critical field $H_{\rm c1}$ ($H_{\rm d}$ above) with no change in the physical domain boundaries.  Then with a further increase of the magnitude of the field all moments progressively cant towards the field giving rise to the observed linear $M(H)$ behavior up to the critical field $H^{\rm c}_{ab}$ noted above.  Our fits describe the experimental $M(H)$ isotherms at $T=2$~K for $H\parallel [1,0,0]$ and $H\parallel [1,1,0]$ with $H\leq H_{\rm c1}$ rather well, where the $H_{\rm c1}$ values for the two field directions are quite different. 

The experimental details are given in Sec.~\ref{Sec:ExpDet}.  The experimental results are presented in Sec.~\ref{Sec:ExpResults}, including magnetic susceptibility $\chi(T)$ data in Sec.~\ref{Sec:MagSus} and magnetization versus field $M(H)$ isotherms in Sec.~\ref{Sec:IsoMag}.  Theoretical fits to the experimental $M(H)$ data at \mbox{$T=2$~K} are given in Sec.~\ref{Fits}.

\section{\label{Sec:ExpDet} Experimental Details}

\eg\ single crystals were grown using an ${\rm EuGa_9}$ self-flux.  The high purity elements (Eu metal from Ames Laboratory and 99.99999\%-pure Ga from Alfa Aesar) were loaded in an alumina crucible and sealed in a silica tube.  The ampule was then heated to 750~$^{\circ}$C at a rate of 100~$^{\circ}$C/h and held for 12~h.  Then it was slowly cooled to 400~$^{\circ}$C at a rate of 2~$^{\circ}$C/h.  The crystals were obtained after removing the flux using a centrifuge.  The sample homogeneity and chemical composition were confirmed using a JEOL scanning electron microscope (SEM) equipped with an EDS (energy-dispersive x-ray spectroscopy) analyzer.  The EDS measurements yielded a composition EuGa$_{4.04(2)}$, close to the stoichiomentric composition.  Magnetic measurements were carried out using a Magnetic-Properties-Measurement System (MPMS) from  Quantum Design, Inc.  We use cgs magnetic units throughout, where $1~{\rm T}  = 10^4$~Oe.

\section{\label{Sec:ExpResults} Experimental Results}

\subsection{\label{Sec:MagSus} Magnetic Susceptibility}

\begin{figure}
\includegraphics [width=3.3in]{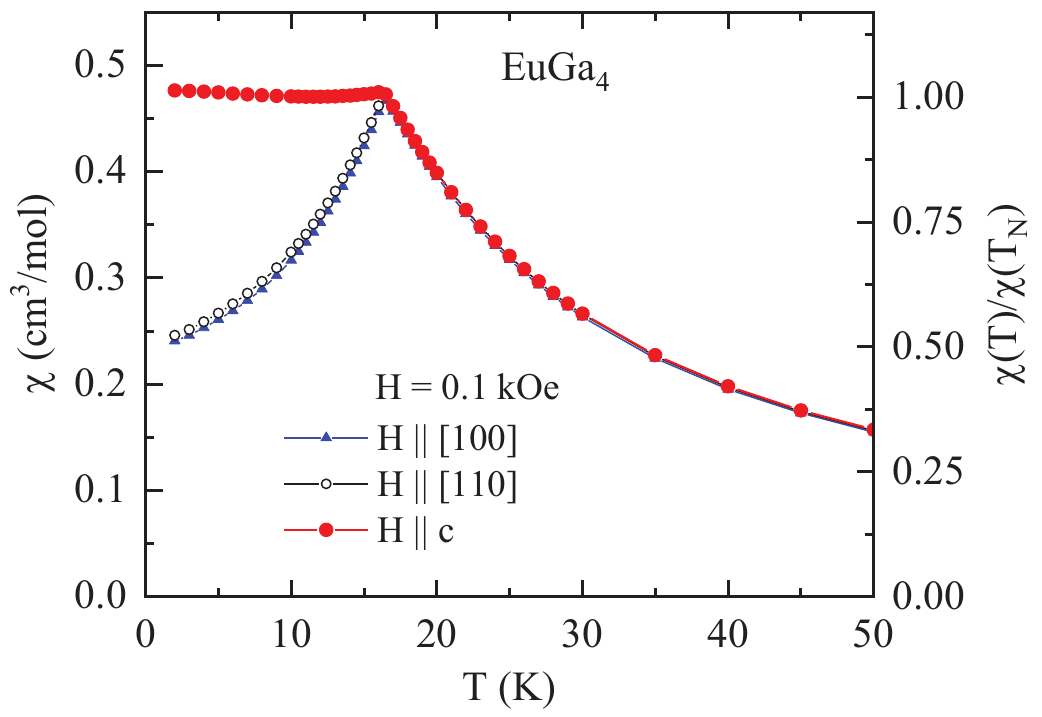}
\caption {Left ordinate: Temperature dependence of the magnetic susceptibility $\chi (T)$ of \eg\ for $H = 0.1$~kOe with \mbox{$H \parallel ab \parallel [1,0,0]$} (filled blue triangles), $H \parallel ab \parallel [1,1,0]$ (open black circles), and $H \parallel c$ (filled red circles). The corresponding susceptibility ratio $\chi (T)/\chi (T_{\rm N})$ is shown on the right ordinate.}
\label{Fig_M-T}
\end{figure}

\begin{figure*}
\includegraphics [width=7in]{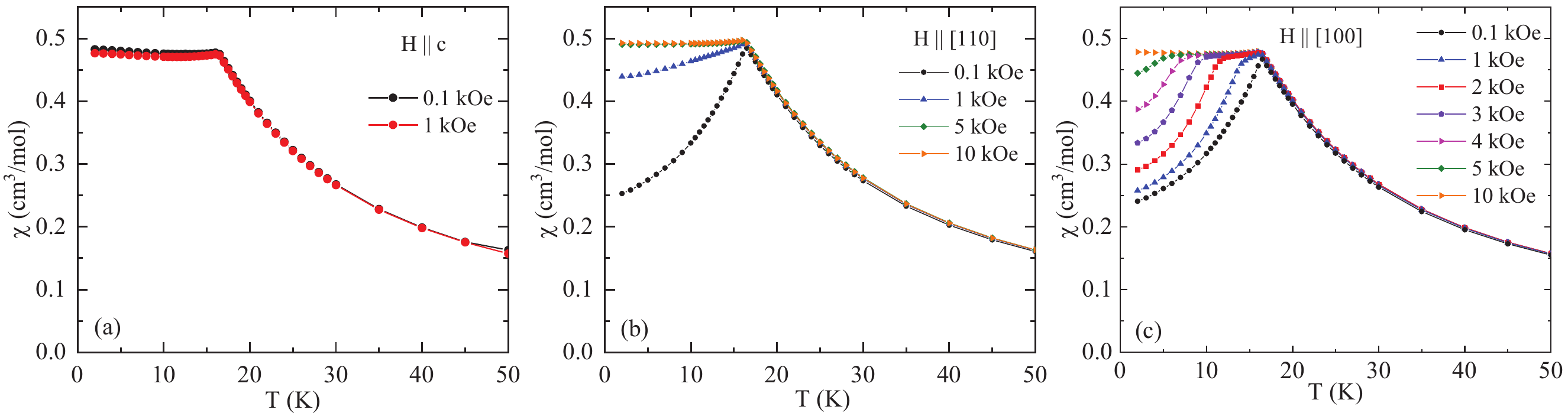}
\caption {(a) Out-of-plane Magnetic susceptibility $\chi_c(T)$ ($H \parallel c$) of \eg\ for different applied magnetic fields. In-plane magnetic susceptibility for different magnetic fields when (b) $H \parallel ab \parallel [1,1,0]$ and (c) $H \parallel ab \parallel [1,0,0]$. The field responses in these two in-plane symmetry directions are significantly different.}
\label{Fig_M-T_diff_fields}
\end{figure*}

The $\chi(T)$ data for \eg\ obtained with in-plane \mbox{($\chi_{ab}, H \parallel ab$)} and out-of-plane ($\chi_{c}, H \parallel c$) magnetic fields $H = 0.1$~kOe are shown in Fig.~\ref{Fig_M-T}, where $\chi_{ab}$ is measured for the two symmetry directions $H \parallel [1,0,0]$ and $H \parallel [1,1,0]$.  The data in the figure indicates that $\chi$ is nearly isotropic in the $ab$~plane.  As seen from the figure, \eg\ undergoes an AFM transition at $T_{\rm N} = 16.4(2)$~K, similar to values reported earlier~\cite{Nakamura2014, Nakamura2015, Nakamura2013, Zhu2022, Kawasaki2016, Yogi2013}.  The $\chi_{c}$ is found to be independent of $T$ for $T \leq T_{\rm N}$, indicating that the moments are aligned perpendicular to the $c$~axis.  This is consistent with the $\chi_{ab}$ data that decrease with decreasing~$T$ with \mbox{$\chi_{ab} (2~{\rm K})/\chi(T_{\rm N})\approx 0.5$}.   According to molecular-field-theory (MFT)~\cite{Johnston2012, Johnston2015} for a $c$-axis helix of identical crystallographically-equivalent Heisenberg spins, one has
\be
\label{Eq:kd}
\frac{\chi_{ab}(T=0)}{\chi_{ab}(T_{\rm N})}=\frac{1}{2[1+2~{\rm cos}(kd)+2~{\rm cos}^2(kd)]},
\ee
where $k$ is the magnitude of the $c$-axis AFM propagation vector, $d$ is the distance along the $c$~axis between the FM layers of spins, and hence $kd$ is the turn angle between the adjacent layers of spins.

The ratio on the left side of Eq.~(\ref{Eq:kd}) for ${\bf H}\parallel[1,0,0]$ at $T=2$~K was previously found to be $\approx 0.26/0.51 \approx 0.51$~\cite{Nakamura2013}.  According to Fig.~\ref{Fig_M-T}, as noted above we find the similar value
\bea
\frac{\chi_{[1,0,0]}(T = 2\,{\rm K})}{\chi(T_{\rm N})} \approx \frac{0.24}{0.48}\approx \frac{1}{2}.
\eea
Using this value of \mbox{$\chi_{ab} (2~{\rm K})/\chi_{ab}(T_{\rm N})$}, Eq.~(\ref{Eq:kd}) yields the turn angle between the moment directions in adjacent Eu layers to be $kd=180^\circ$, indicating that the AFM structure is \mbox{A-type}, in agreement with the earlier neutron-diffraction solution of the magnetic structure of \eg~\cite{Kawasaki2016}.

\begin{figure}
\includegraphics [width=3in]{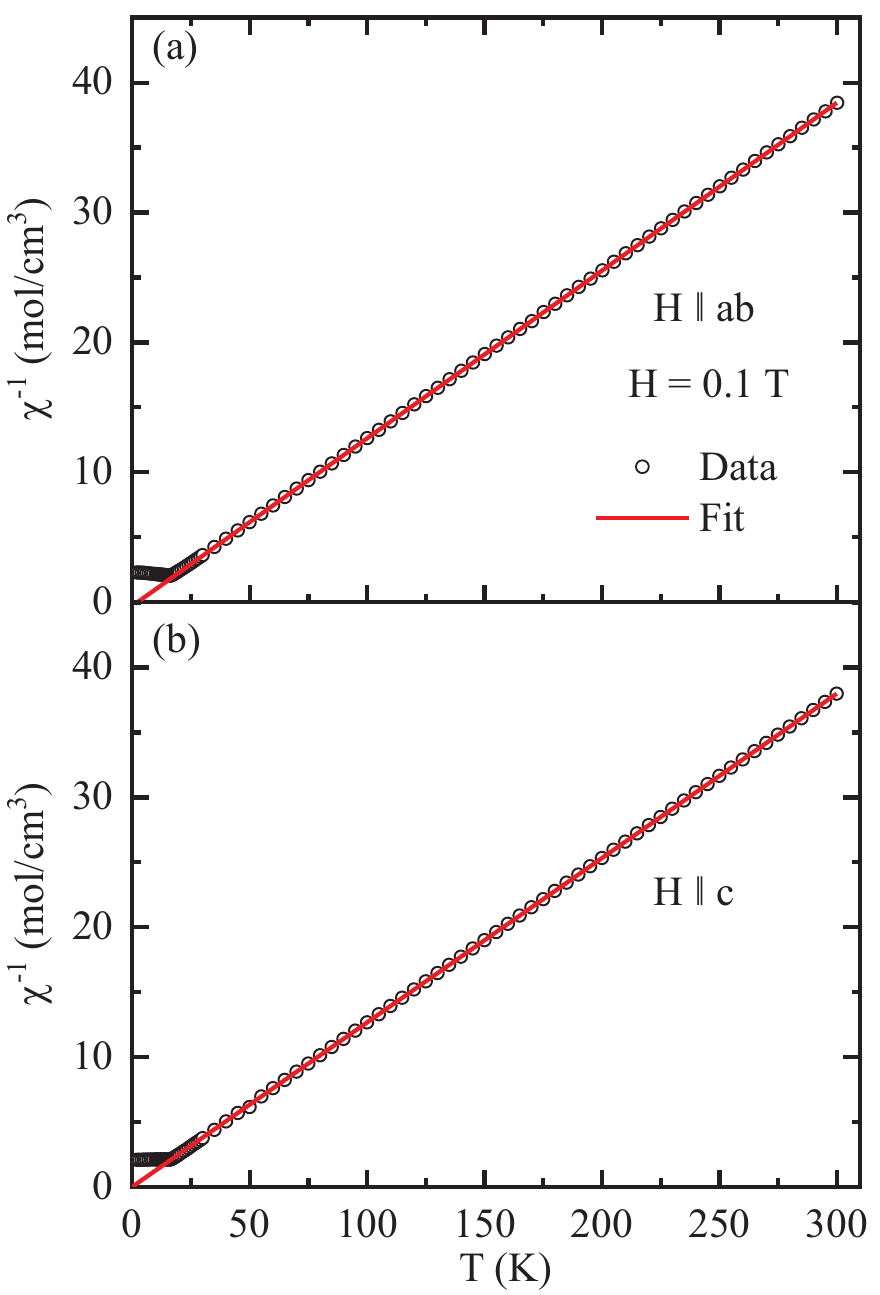}
\caption {Inverse magnetic susceptibility $\chi^{-1}(T)$ of \eg\ with $H= 0.1$~T for (a)~$H \parallel ab$ and (b)~$H\parallel c$.}
\label{Fig_Chi-1-T}
\end{figure}

The inverse molar magnetic susceptibility $1/\chi$ for $H=0.1$~T is plotted versus~$T$ for $H\parallel [1,0,0]$ in Fig.~\ref{Fig_Chi-1-T}(a) and $H\parallel [0,0,1]$ in Fig.~\ref{Fig_Chi-1-T}(b).  Both plots are linear above $T_{\rm N}$ and are described well by the inverse of the modified Curie-Weiss law
\bea
\chi(T) = \chi_0+\frac{C}{T-\theta},
\eea
where $\chi_0$ is the $T$-independent contribution, $C$ is the molar Curie constant and $\theta$ is the Weiss temperature.  The fits yield the values of these variables in Table~\ref{Tab.chidata}. The magnitudes of the diamagnetic $\chi_0$ values are of the order expected for the diamagnetic core contributions but are very small relative to the $\chi$ values of the Eu$^{2+}$ moments.  The listed effective moments $\mu_{\rm eff}$ for the two field directions are close to the theoretical value of 7.94~$\mu_{\rm B}$/Eu$^{2+}$ for $g=2$ and  $S=7/2$.  The Weiss temperatures are positive, consistent with the A-type AFM structure in which FM planes of Eu spins are stacked antiferromagnetically along the $c$~axis.  However, they are not close to the value of $T_{\rm N}$, indicating that the AFM interactions between the Eu spins in adjacent layers perpendicular to the $c$~axis are also significant.

\begin{table}
\caption{\label{Tab.chidata} The fitted parameters to the inverse susceptibility data in Fig.~\ref{Fig_Chi-1-T}, including the $T$--independent contribution to the susceptibility $\chi_0$, molar Curie constant $C_\alpha$ for $\alpha = ab,\,c$ directions, effective moment per Eu spin $\mu_{\rm{eff}} \approx \sqrt{8C_\alpha}$ and the Weiss temperature $\theta_\alpha$.}
\begin{ruledtabular}
\begin{tabular}{ccccc}	
Field & $\chi_0$ 				& $C_{\alpha}$ 		    &  $\mu_{\rm eff\alpha}$ 	& $\theta_\alpha$ \\
direction	 	& $\rm{\left(10^{-5}~\frac{cm^3}{mol}\right)}$	 & $\rm{\left(\frac{cm^3 K}{mol}\right)}$    & $\rm{\left(\frac{\mu_B}{mol}\right)}$& (K)  \\
\hline
H$\parallel ab$ 		& -1.7(5)		&  	7.76(1)	&	7.88(1)		& 2.27(6)		\\
H$\parallel c$ 			& $-3.5(3)$ 		&  	7.86(1)	&	7.93(1)	& 0.5(1)	\\
\end{tabular}
\end{ruledtabular}
\end{table}

The magnetic-field dependences of $\chi(T)$ are shown in Fig.~\ref{Fig_M-T_diff_fields} for (a) $H \parallel c$, (b) $H \parallel [1,1,0]$ and (c) $H \parallel [1,0,0]$.  No change in $\chi_c(T)$ is observed between $H = 0.1$ and 1~kOe.  However, a significant variation of $\chi_{ab}(T)$ is observed in this field region.  Interestingly, the field evolution of $\chi_{ab}(T)$ at low fields is quite different when the applied field is applied along the $ab$~plane [1,0,0] and [1,1,0] directions. The critical fields at which the moments become aligned with the applied field are at much higher fields $H_{ab}^{\rm c} = 71$~kOe and $H_{c}^{\rm c} = 72$~kOe for $H \parallel ab$ and $H \parallel c$, respectively~\cite{Nakamura2013}, indicating a very small magnetic anisotropy between these two field directions as expected for Eu$^{2+}$ moments with $S=7/2$ and $L=0$. 

The $\chi_{ab}(T)$ in Fig.~\ref{Fig_Chi-1-T}(b) for $T \leq T_{\rm N}$ strongly increases between applied fields $H = 0.1$ and 1~kOe applied along the [1,1,0] direction and at higher fields it becomes independent of $T$\@.  On the other hand, only a gradual  increase in $\chi_{ab}(T)$ with increasing~$H$ is observed for $H \parallel [1,0,0]$ in Fig.~\ref{Fig_Chi-1-T}(c) in the field range ${\rm 0.1~kOe} \leq H \lesssim {\rm 6~kOe}$.  Moreover, a $T$-independent region of $M(H)$ is observed for $H = 1$~kOe for $H \parallel [1,0,0]$ and the temperature range of that plateau increases with increasing $H$.  Finally, $\chi(T)$ for both $H \parallel [1,0,0]$ and $H \parallel [1,1,0]$ in the AFM state below $T_{\rm N}$ becomes independent of $T$ for $H = 10$~kOe.  We show in Sec.~\ref{Fits} below that the different low-field $M(H)$ behavior of $\chi_{ab}(H)$ for $H \parallel [1,0,0]$ and $H \parallel [1,1,0]$ in \eg\  is due to AFM domain formation arising from the fourfold tetragonal $c$-axis rotational symmetry.  Similar effects were  found previously in trigonal Eu-based compounds with threefold rotational symmetry about the $c$~axis~\cite{Pakhira2020, Pakhira2021, Pakhira2022b, Pakhira2021a, Pakhira2022}.  

\subsection{\label{Sec:IsoMag} Magnetization Isotherms}

\begin{figure}
\includegraphics [width=3.3in]{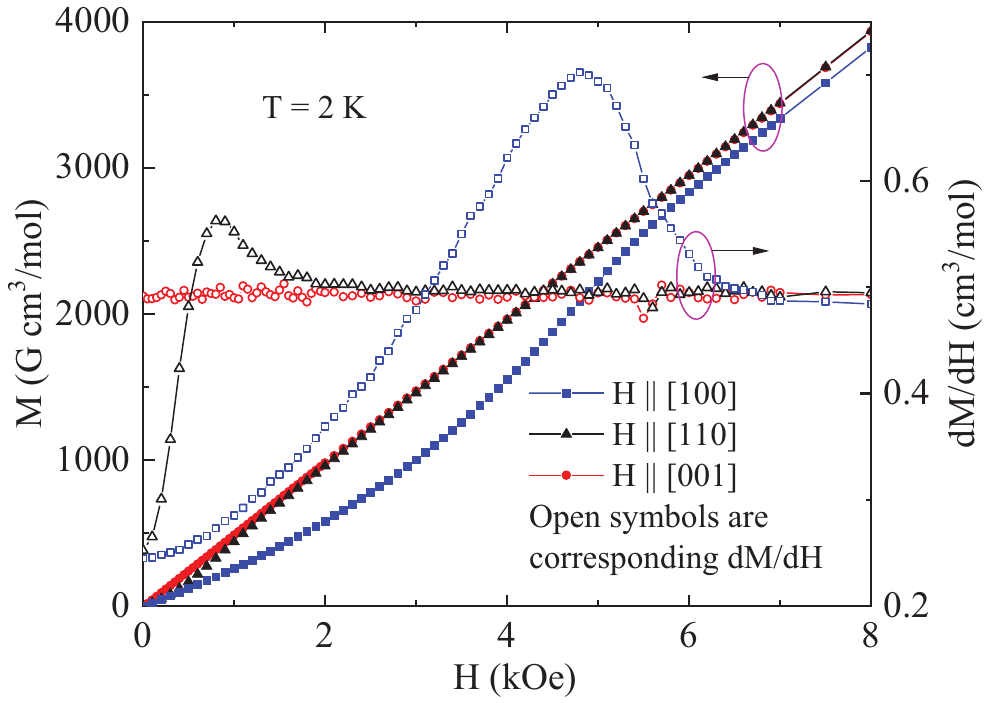}
\caption {Low-field $M(H)$ data measured at $T = 2$~K for $H \parallel [1,0,0]$, $H \parallel [1,1,0]$, and $H \parallel [0,0,1]$ (left ordinate). The corresponding field derivatives are also plotted (right ordinate). Although $M_c(H)$ for $H \parallel [0,0,1]$ is linear, distinct nonlinearities in $M_{ab}(H)$ for $H \parallel [1,0,0]$ and [1,1,0] are observed in this low-field region.  }
\label{Fig_M-H_2 K}
\end{figure}

In order to provide further insight into the field-dependent evolution of the magnetic behavior at $T<T_{\rm N}$, $M(H)$ isotherm data were obtained that emphasize the low-field region of interest.  As can be seen in Fig.~\ref{Fig_M-H_2 K}, the $M(H)$ behavior measured at $T = 2$~K for $H \parallel c$ is linear.  In accordance with the magnetic susceptibility measurements, a clear nonlinear response in $M(H)$ is observed for both $H \parallel [1,0,0]$ and $H \parallel [1,1,0]$.  This is clearly reflected in the $dM/dH$ data, where $dM/dH$ initially increases rapidly with increasing $H$ and exhibits peaks at the critical fields $H_{c1, [1,0,0]} \approx 4.8$~kOe and $H_{c1, [110]} \approx 0.85$~kOe, followed eventually by an $H$-independent behavior for $H > H_{c1}$.  The difference in the low-field $M(H)$ behavior for different in-plane symmetry directions can be explained by the rotation of the moments in $ab$-plane AFM domains as discussed in detail below.  

\begin{figure*}
\includegraphics [width=6.in]{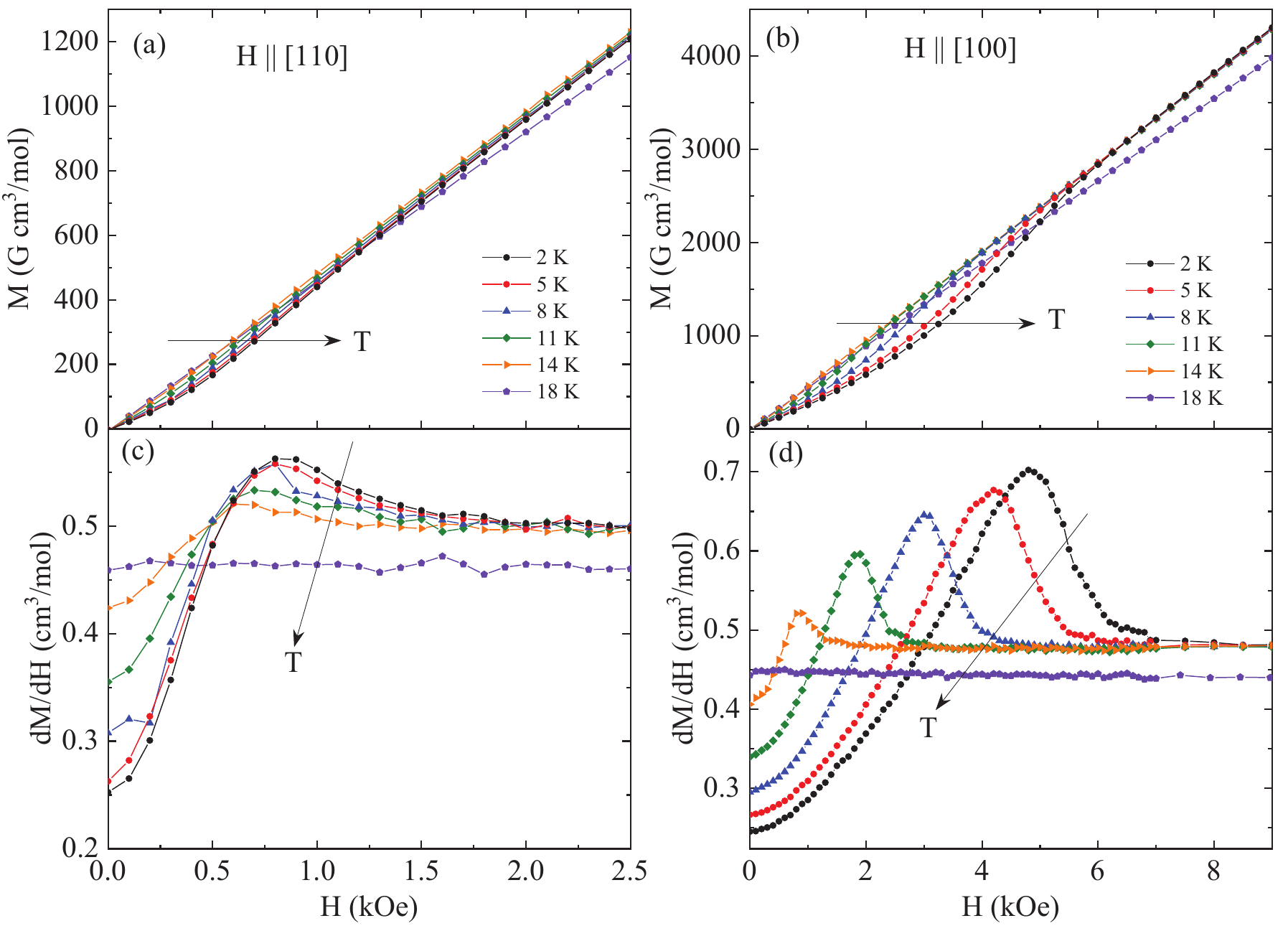}
\caption {In-plane $M(H)$ data measured at different temperatures for (a) $H \parallel [1,1,0]$ and (b) $H \parallel [1,0,0]$. The respective $dM(H)/dH$ versus~$H$ data are shown in~(c) and~(d). The data for $T=18$~K are about 2~K above $T_{\rm N}$.}
\label{Fig_M-H_diff_temp}
\end{figure*}

The $M(H)$ data measured at different temperatures for $H \parallel [1,1,0]$ are shown in Fig.~\ref{Fig_M-H_diff_temp}(a) and the corresponding $dM/dH$ versus~$H$ data are plotted in Fig.~\ref{Fig_M-H_diff_temp}(c). As seen in the latter figure, $H_{c1}$ slightly shifts to lower fields with increasing temperature below $T_{\rm N}$, with \mbox{$H_{c1, [1,1,0]} = 0.85$~kOe} at $T = 2$~K decreasing to 0.6~kOe at $T = 14$~K\@.  The $M(H)$ behavior is linear for \mbox{$T > T_{\rm N}$.}  The $T$ dependences of $M(H)$ and $dM/dH(H)$ for \mbox{$H \parallel [1,0,0]$} are shown in Figs.~\ref{Fig_M-H_diff_temp}(b) and \ref{Fig_M-H_diff_temp}(d), respectively.  Here, the nonlinearity in $M(H)$ at $T = 2$~K persists up to $H \approx 0.8$~T, which is much larger than that observed for $H \parallel [1,1,0]$.  The $dM/dH$ for this field direction shows a maximum at $H_{c1, [1,0,0]} = 4.8$~kOe at $T = 2$~K\@.  This critical field is significantly reduced to \mbox{$H_{c1, [1,0,0]} = 0.85(5)$~kOe} at $T = 14$~K\@.  The striking difference observed in the $M(H)$ and corresponding $dM/dH$ behavior between the $H \parallel [1,1,0]$ and $H \parallel [1,0,0]$ directions indicates the presence of significant in-plane magnetic anisotropy.  This anisotropy is associated with the magnetic-field-induced moment reorientation in the AFM domains discussed below in Sec,~\ref{Fits}.

\begin{figure}
\includegraphics [width=3.3in]{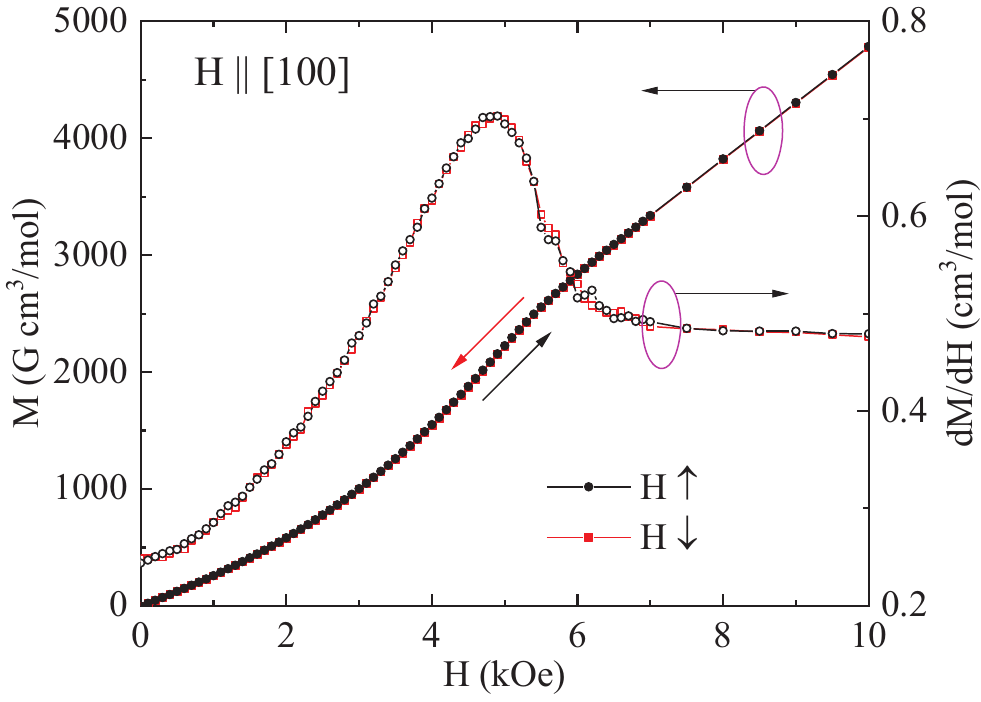}
\caption {In-plane field-cooled $M(H)$ behavior for $H \parallel [1,0,0]$ in the hysteresis $H$ cycle 10~kOe $\rightarrow$ 0 $\rightarrow$ 10~kOe. No hysteretic behavior is observed. The corresponding $dM(H)/dH$ behavior is also shown in the right ordinate.}
\label{Fig_M-H_2 K_hysteresis}
\end{figure}

We tested the reversibility of the nonlinear $M(H)$ and $dM/dH$ at low fields upon heating and cooling for $H \parallel [1,0,0]$.  The crystal was initially cooled to $T = 2$~K under a magnetic field $H = 5$~T\@.  After $T$ stabilization, $M(H)$ was measured in the hysteresis $H$ cycle 10~kOe $\rightarrow$ 0~kOe $\rightarrow$ 10~kOe, as shown in Fig.~\ref{Fig_M-H_2 K_hysteresis}.  No magnetic hysteresis was observed, indicating that the low-field-induced $M(H)$ nonlinearity is reversible.

Similar $M_{ab}(H)$ behavior was observed for the Eu-based trigonal compounds \emb\ and \ems~\cite{Pakhira2020, Pakhira2021, Pakhira2021a, Pakhira2022} and we successfully modeled those results~\cite{Pakhira2022b} using an approach similar to that used below to model the low-field $M(H)$ data for \eg.

\section{\label{Fits} Fits to the Experimental M(H) Data}

\subsection{Theory}

\begin{figure}[h]
\includegraphics [width=3.3in]{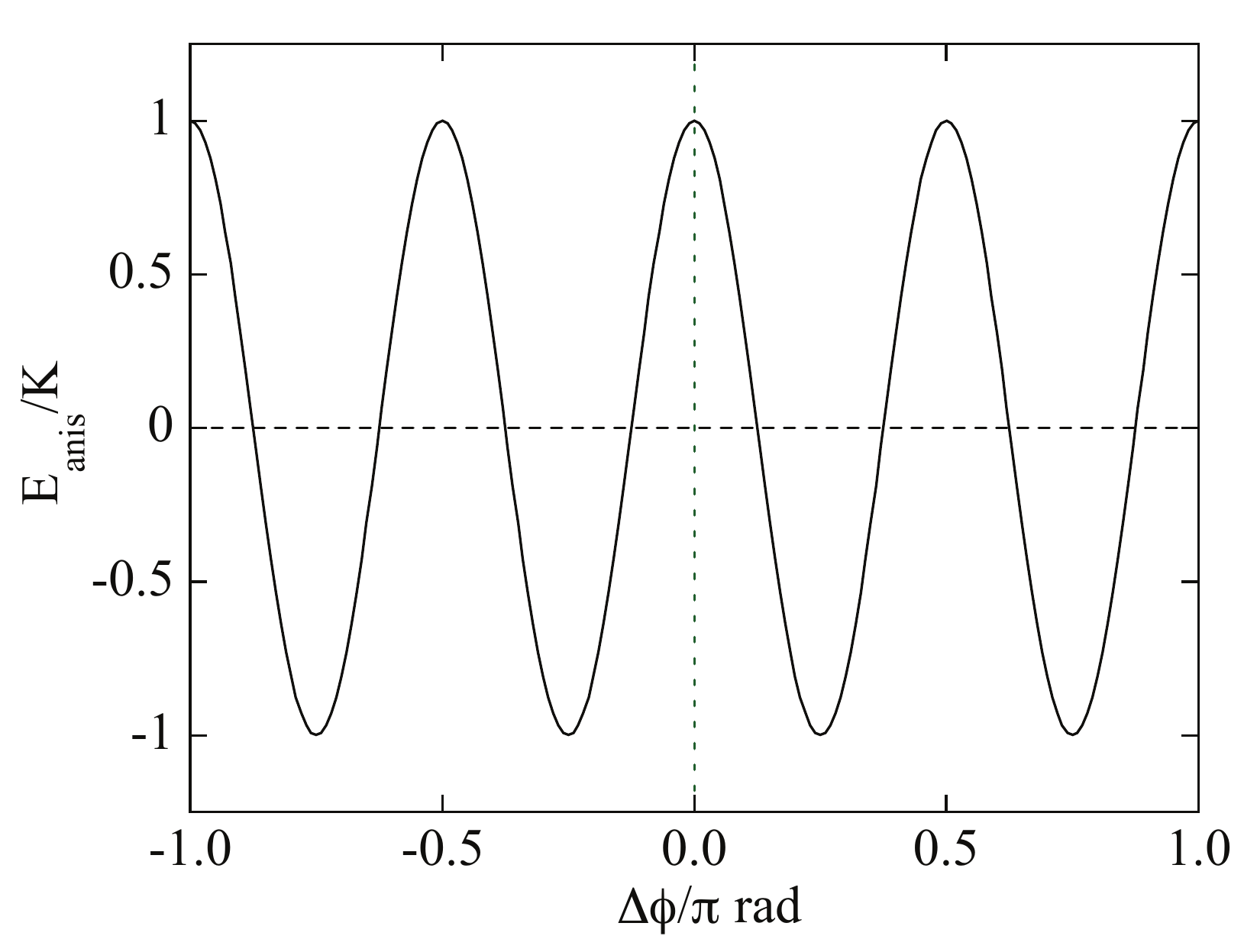}
\caption {Fourfold $ab$-plane rotational anisotropy energy normalized by the anisotropy constant $K$, $E_{\rm anis}/K$, versus the $ab$-plane tilt angle $\Delta\phi/\pi$\,rad of a moment in tetragonal \eg.}
\label{Fig_4foldAnis}
\end{figure}

We write the fourfold rotational magnetic anisotropy energy $E_{\rm anis}$ for the ferromagnetic $ab$-plane layers in tetragonal \eg\ versus the azimuthal angle $\phi$ of the ferromagnetic moments in that layer as
\bea
E_{\rm anis} = K_4 \cos(4\phi),
\label{Eq:Eanis}
\eea
where $K_4>0$ is the fourfold $ab$~plane anisotropy constant and $\phi$ is the angle of the moments with respect to the $x$ axis defined in Fig.~\ref{Figure_domain} below.  A plot of $E_{\rm anis}/K_4$ versus $\phi$ is shown in Fig.~\ref{Fig_4foldAnis}.  The anisotropy-energy minima occur at $\phi = \pm \pi/4$ and $\pm 3\pi/4$~rad.

\begin{figure}
\includegraphics [width=2.5in]{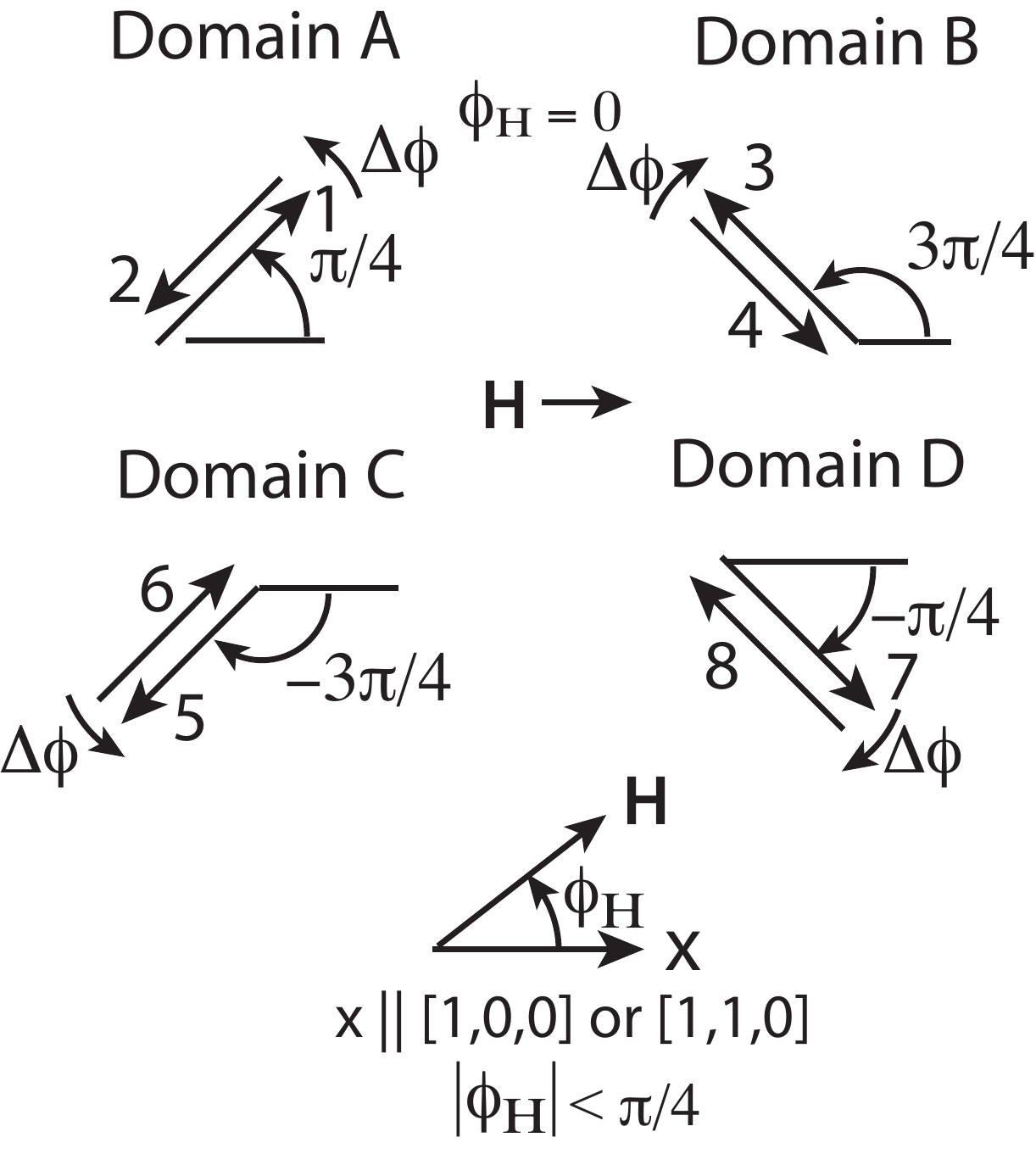}
\includegraphics [width=1.55in]{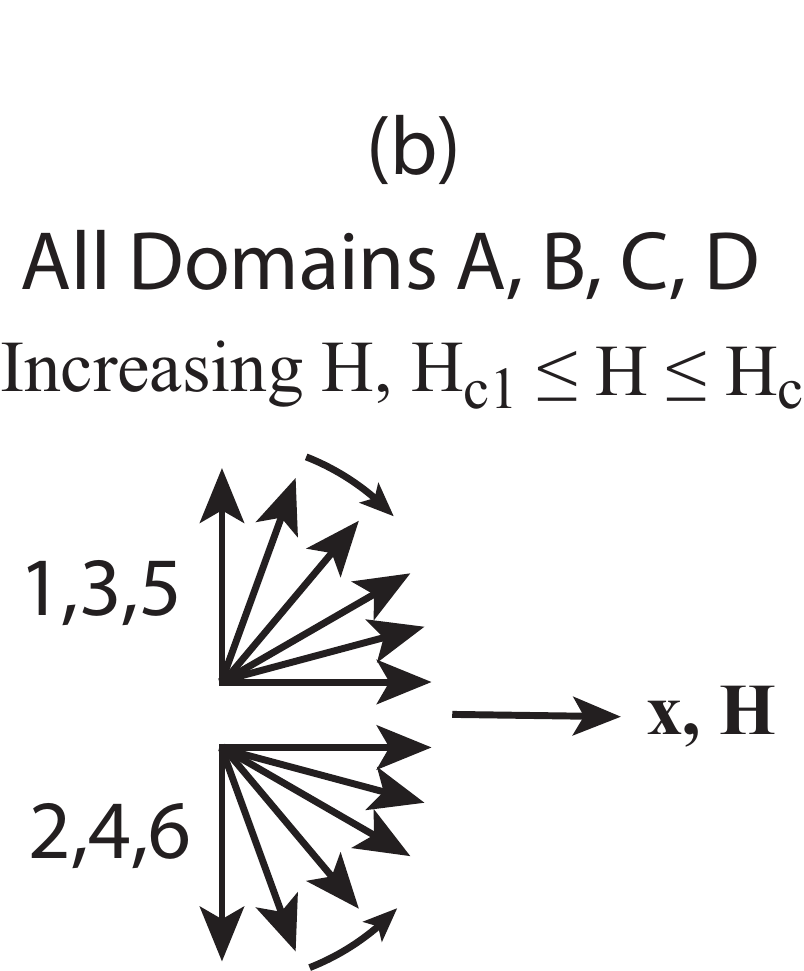}
\caption {(a)~Schematic diagram of the nearly-locked moment orientations in the $ab$-plane of adjacent antiparallel layers of moments along the $c$~axis in the four collinear A-type AFM domains A, B, C, and D and their magnetic field evolution with increasing $x$-axis field $H$ at low fields $H \leq H_{\rm c1}$ shown by arrows.  (b)~For $H_x>H_{\rm c1}$, all pairs of moments in adjacent ferromagnetically-aligned layers increasingly cant by the same amount towards the increasing field as shown, until at the critical field $H_{\rm c}$ all moments are aligned with the field.}
\label{Figure_domain}
\end{figure}

In order to model the anomalous low-field $M_{ab}(H)$ behavior for \eg\ in Figs.~\ref{Fig_M-H_2 K}--\ref{Fig_M-H_2 K_hysteresis},  we propose that the magnetic structure in $H=0$ contains four equally-populated domains A, B, C, D of ferromagnetically-aligned moments in the \mbox{A-type} AFM structure illustrated in Fig.~\ref{Figure_domain}(a) as required by the tetragonal lattice symmetry.  As shown in Fig.~\ref{Figure_domain}(a), each of the four domains contains moments that are ferromagnetically-aligned in every-other $ab$~plane and the moments in adjacent layers along the $c$~axis are aligned at $180^\circ$ with respect to the former moments, as required for an A-type AFM structure.  We assume that within each physical AFM domain, the applied field $H_x$ rotates the moments only in the $ab$~plane and does not cause domain-wall motion.   The former assumption is justified because in a body-centered-tetragonal lattice, the magnetic-dipole interaction favors ferromagnetic moment alignment in the $ab$~plane rather than along the $c$~axis~\cite{Johnston2016}.  As noted at the bottom of Fig.~\ref{Figure_domain}(a), the $x$ direction of the applied field can be aligned along either the crystallographic [1,0,0] or [1,1,0] directions which are expected to have different anisotropy energies.

In $H_x = 0$, the angles of moments 1, 3, 5, and 7 with respect to the $x$ axis in the respective ferromagnetically-aligned layer are in energy minima according to Fig.~\ref{Fig_4foldAnis}.  Similarly, moments 2, 4, 6, and 8 in either of the two layers of the A-type AFM structure adjacent to the respective layers containing moments 1, 3, 5, and 7 are also in energy minima.  On application of $H_x$, due to the relationship of the directions of the moments in the different domains in Fig.~\ref{Figure_domain}(a) to each other, the magnitude of the change of the moment angle $\Delta \phi$ is the same for the moments in each domain as shown in the figure.  During this process, the moments in adjacent ferromagnetically-aligned layers retain their 180$^\circ$ alignment due to the AFM exchange interaction between moments in adjacent layers, apart from a very small canting towards the field which gives rise to the observed magnetization.

The moments in each domain eventually rotate to become perpendicular to $H_x$ at a critical field $H_{\rm c1}$ at which a maximum is observed in $dM_{ab}/dH$ in \mbox{Figs.~\ref{Fig_M-H_2 K}--\ref{Fig_M-H_2 K_hysteresis}.}   For $H_x > H_{c1}$, the moments in each domain increasingly cant towards the applied field direction as shown in Fig.~\ref{Figure_domain}(b).  The magnetization saturates when all the moments become parallel to the applied field $H_x$ at a critical field~$H_{\rm c}$.  As noted at the bottom of Fig.~\ref{Figure_domain}(a), it is possible to align {\bf H} at an angle $\phi_{H}\neq0$ with respect to the positive $x$~axis.   As illustrated later, at $T=0$~K this is predicted to result in a first-order transition at $H_{\rm c1}$.

For $0\leq H_x\leq H_{\rm c1}$, from Fig.~\ref{Figure_domain}(a) the angles of the moments in each domain with respect to the positive $x$~axis  are respectively
\bea
\phi_{12} &=& \frac{\pi}{4} + \Delta\phi, \nonumber\\
\phi_{34} &=& \frac{3\pi}{4} - \Delta\phi  \label{Eqs:phiABCD}\\
\phi_{56} &=& -\frac{3\pi}{4} + \Delta\phi, \nonumber\\
\phi_{78} &=& -\frac{\pi}{4} - \Delta\phi, \nonumber\\
\nonumber
\eea
where $0\leq \Delta\phi \leq \pi/4$.  The average anisotropy energy per domain for $0\leq H_x\leq H_{\rm c1}$ obtained using Eqs.~(\ref{Eq:Eanis}) and~(\ref{Eqs:phiABCD}) is 
\bea
E_{\rm anis\,ave}  = -K_4 \cos(4\Delta\phi).
\label{Eq:EanisAve}
\eea

The magnetization component $\mu_\parallel$ of a moment along the axis of a pair of collinear moments aligned antiparallal in domain~$n$ with azimuthal angle $\phi_n$ in Fig.~\ref{Figure_domain}(a) at $T=0$ is zero, whereas the component $\mu_\perp$ of a moment perpendicular to the moment is $\mu_\perp = \chi_\perp H_{x\perp}=  \chi_\perp H_x \sin(\phi_n)$, where $\chi_\perp$ is the magnetic susceptibility per moment in a domain when the field is perpendicular to it.  The angle $\phi_n$ is the angle of domain~$n$ in Fig.~\ref{Figure_domain}(a) with respect to the $x$~axis given by Eqs.~(\ref{Eqs:phiABCD}).  The component of the applied field in the direction perpendicular to the moment is $H_\perp = H_x\sin(\phi_n)$. Therefore the magnetic energy of a moment in domain~$n$ in the regime $0\leq H_x \leq H_{\rm c1}$ is given by
\bse
\label{Eq:Etotal}
\bea
E_{{\rm mag}\,n} &=& \mu_xH_x \nonumber\\
&=& -[\chi_\perp H_x \sin(\phi_n)] [H_x\sin(\phi_n)] \label{Eq:Emag}\\
&=& -\chi_\perp H_x^2\sin^2(\phi_n).\nonumber
\eea
Summing $E_{{\rm mag}\,n}$ over the angles of the moments in the four domains in Eq.~(\ref{Eqs:phiABCD}) and dividing by four gives the average magnetic energy per moment for $0\leq H_x\leq H_{\rm c1}$.  Then using Eq.~(\ref{Eq:EanisAve}) and normalizing the total energy per moment by $K_4$ gives the total average energy per moment as
\bea
E_{\rm ave}/K_4 &=& (E_{\rm anis\,ave} + E_{\rm mag\,ave})/K_4 \label{Eq:EmagAve} \\
&=& - \cos(4\Delta\phi) -\frac{\chi_\perp H_x^2}{2}\big[1+\sin(2\Delta\phi)\big].\nonumber
\eea
\ese

\begin{figure}
\includegraphics [width=3.3in]{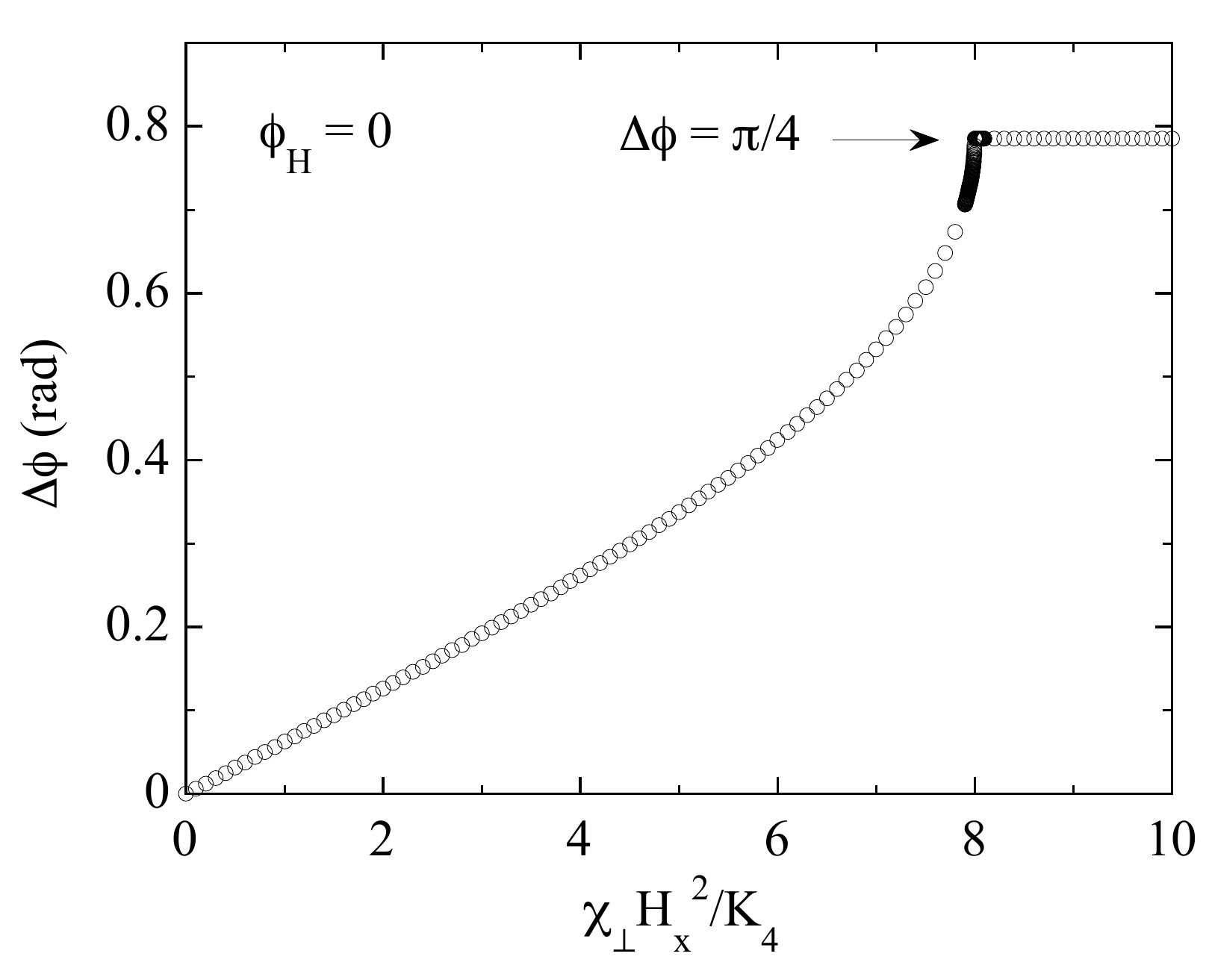}
\caption {The angle $\Delta\phi$ in Fig.~\ref{Figure_domain}(a) versus $\chi_\perp H_x^2/K_4$ for \mbox{$\phi_H=0$~rad} as defined at the bottom of  Fig.~\ref{Figure_domain}(a).}
\label{Fig_Phi1_2VsH_PhiH0}
\end{figure}

Setting the derivative of $E_{\rm ave}/K_4$ with respect to $\Delta\phi$ in Eq.~(\ref{Eq:EmagAve}) equal to zero gives $\Delta\phi$ versus $\chi_\perp H_x^2/K_4$ as plotted in Fig.~\ref{Fig_Phi1_2VsH_PhiH0}. The maximum value  $\Delta\phi = \pi/4$ is the value at which all moments become perpendicular to the applied field $H_x=H_{c1}$, apart from a small canting of all moments towards the field that gives rise to the measured magnetization along the $x$ axis.  At larger fields up to the critical field $H_{\rm c}$, the individual moments cant towards the field and molecular-field theory predicts \mbox{$\chi = \chi_\perp H_x = \chi(T_{\rm N})H_x$} until the critical field $H_{\rm c}$ is reached at which all the moments are aligned with the applied field.

Figure~\ref{Fig_Phi1_2VsH_PhiH0} shows the the value of~$\chi_\perp H_x^2/K_4$ at the critical field $H_{c1}$ for which $\Delta\phi=\pi/4$~rad is given by
\bse
\label{Eqs:chiPerpHc12K}
\bea
\frac{\chi_\perp H_{\rm c1}^2}{K_4} = 8,
\label{Eq:Hc1Def}
\eea
where the factor of 8 appears to be exact. This equation gives the value of the anisotropy constant~$K_4$ in terms of measurable quantities as
\bea
K = \frac{\chi_\perp H_{\rm c1}^2}{8},
\label{Eq:K}
\eea

Since $K$ is normalized to a single moment, whereas the measured $\chi_\perp$ is normalized to a mole of moments, this difference can be taken into account by writing Eq.~(\ref{Eq:K}) as
\bea
K_4 = \frac{\chi_\perp H_{\rm c1}^2}{8N_{\rm A}},
\label{Eq:K2}
\eea
\ese
where $N_{\rm A}$ is Avogadro's number.  For \eg, the measured values are \mbox{$\chi_{ab}(T_{\rm N})=0.48~{\rm cm^3/mol}$} and \mbox{$H_{\rm c1}\approx 4.8$~kOe} for \mbox{$H_x\parallel[1,0,0]$,} yielding
\bea
K_4 = 1.43\times10^{-3}~{\rm \frac{meV}{Eu~atom}}.
\eea
This value is of order 100 times larger than the values of the \mbox{threefold} anisotropy constants $K_3 = 6.5\times10^{-5}$ and $1.8\times10^{-5}$ ~$\rm{\frac{meV}{Eu~atom}}$ obtained for trigonal  \emb\ and \ems, respectively~\cite{Pakhira2022b}.

\subsection{Fits of the M(H) data at T = 2~K by theory}

The magnetization $M_x$ per mole of Eu moments versus magnetic field $H_x$ is calculated from
\bea
M_x(H) &=& \frac{\chi_\perp H_x}{2}\left[\sin^2\left(\frac{\pi}{4} + \Delta\phi\right)+\sin^2\left(\frac{3\pi}{4} - \Delta\phi\right)\right] \nonumber\\
&=& \frac{\chi_\perp H_x}{2}\left[1+\sin(2\Delta\phi)\right],
\label{Eq:MxH}
\eea
where $\chi_\perp$ is equal to the molar magnetic susceptibility at $T_{\rm N}$ and $\Delta\phi$ is given by the data in Fig.~\ref{Fig_Phi1_2VsH_PhiH0}.

\begin{figure}
\includegraphics [width=3.3in]{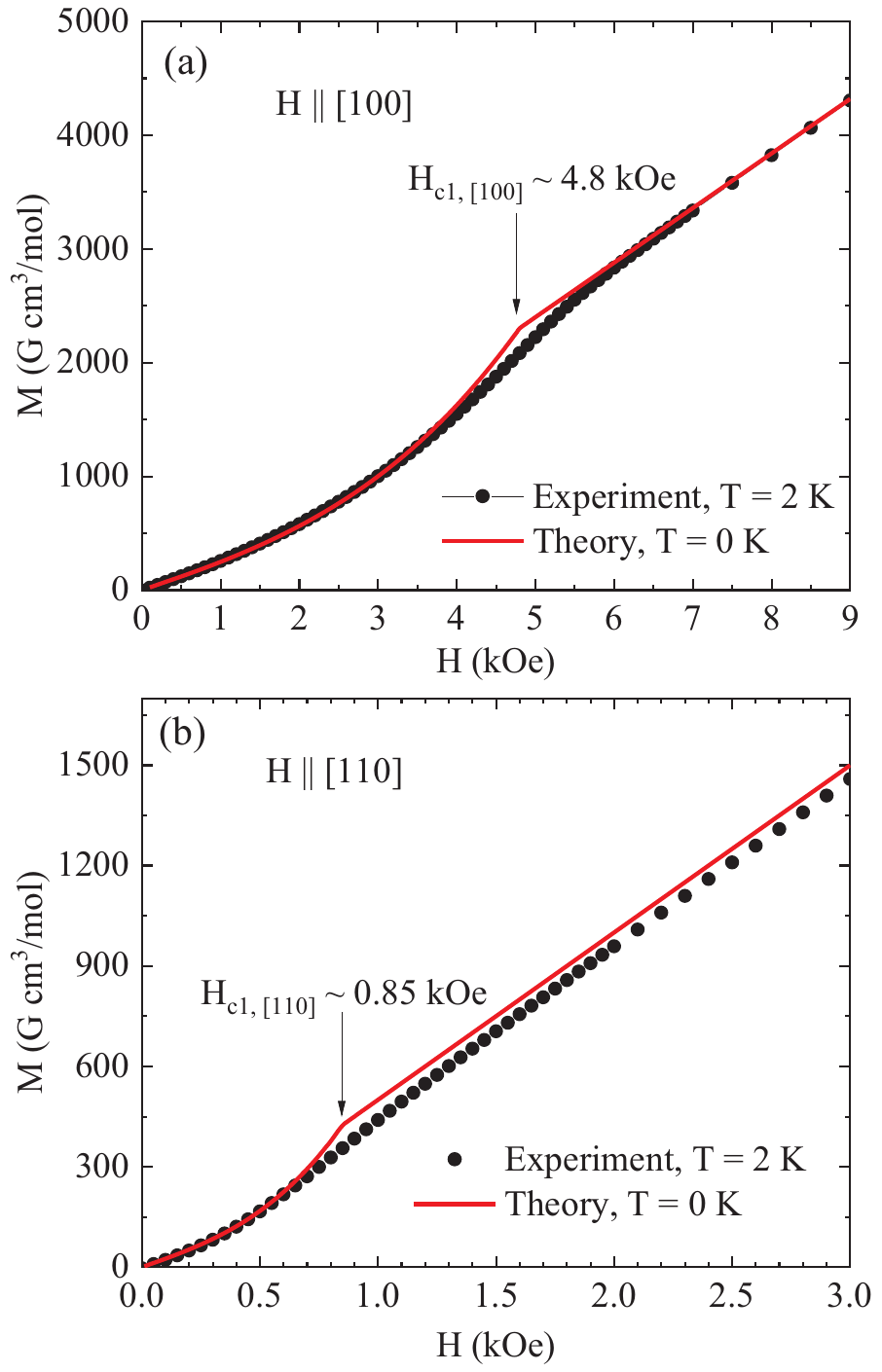}
\caption {The experimental $ab$-plane $M(H)$ behavior for two applied field directions (a) $H \parallel [1,0,0]$ and (b) $H \parallel [110]$ measured at $T = 2$~K along with theoretical predictions for $T = 0$~K with $H_{c1} \approx 4.8$~kOe and 0.85~kOe, respectively, as shown. }
\label{Fig_M-H_fit}
\end{figure}

The experimental $M_{ab}(H)$ data for $H_x \parallel [1,0,0]$ and $H_x \parallel [1,1,0]$ directions along with the calculated theoretical $M_x(H)$ behavior using Eq.~(\ref{Eq:MxH}) are shown in Figs.~\ref{Fig_M-H_fit}(a) and \ref{Fig_M-H_fit}(b), respectively.  The value of $H_{\rm c1}$ is seen to be quite different for the two field directions.  The theory reproduces the experimental data rather well at low and high fields, but deviates somewaht from the data for $H\gtrsim H_{\rm c1}$.  The reason for this discrepancy is not clear at present but may be associated with the fact that the theoretical calculations are done for $T = 0$~K, whereas the observed $M_x(H)$ data were obtained at $T=2$~K\@.   A larger discrepancy between the theoretical and experimental data taken at $T = 1.8$~K was also observed earlier for trigonal \emb\ and \ems, where the measurement temperatures were  $T \approx 0.27~T_{\rm N}$ and $T \approx 0.23~T_{\rm N}$, respectively.  The discrepancy is smaller for \eg\ with measurement temperature $T \approx 0.13~T_{\rm N}$.  This is expected since the $T = 0$~K theoretical predictions are expected to most accurately agree with the  $M(H)$ data when the data are measured at $T\ll T_{\rm N}$.

We have also calculated the $M(H)$ behavior when $H_x$ is not along the $x$ axis parallel to the [1,0,0] or [1,1,0] direction, but is in a direction in the $ab$ plane where $\bf H$ is at an arbitrary  positive angle $\phi_H < \pi/4$ with respect to the $+x$ axis as indicated at the bottom of Fig.~\ref{Figure_domain}(a).  In this case there are effectively two domains A and B, because $\phi_{\rm C} = -\phi_1$ and $\phi_{\rm D} = -\phi_2$. We therefore minimize the energy only with respect to $\phi_1$ of Domain~A and $\phi_2$ of Domain~B\@. The angles of the two domains for $0\leq H_x\leq H_{\rm c1}$ with respect to $\phi_H$ are
\bea
\phi_1 - \phi_H &=& \frac{\pi}{4} + \Delta\phi_1 \quad (0 \leq \Delta\phi_1 \leq \pi/4 + \phi_H), \label{Eqs:phiAB}\\
\phi_2 - \phi_H &=& -\frac{\pi}{4} - \Delta\phi_2.  \quad (0 \leq \Delta\phi_1 \leq \pi/4 - \phi_H).\nonumber
\eea
The average anisotropy energy associated with the two domains is
\bea
E_{\rm anis\,ave} &=& -\frac{K}{2}[\cos(4\phi_1) +  \cos(4\phi_2)].\label{Eq:EanisAve2}
\eea
\noindent The average magnetic energy in the regime $0\leq H_x \leq H_{\rm c1}$ is given by
\bea
E_{\rm mag\ ave} &=& -\frac{\chi_\perp H_x^2}{2}[\sin^2(\phi_1-\phi_H)+\sin^2(\phi_2-\phi_H)] \nonumber \\
&=& -\frac{\chi_\perp H_x^2}{4}\big[2+\sin[2(\phi_1-\phi_H)]\label{Eq:Emag2}\\
&&+\sin[2(\phi_2-\phi_H)]\big].\nonumber
\eea

\begin{figure}[t]
\includegraphics [width=3.3in]{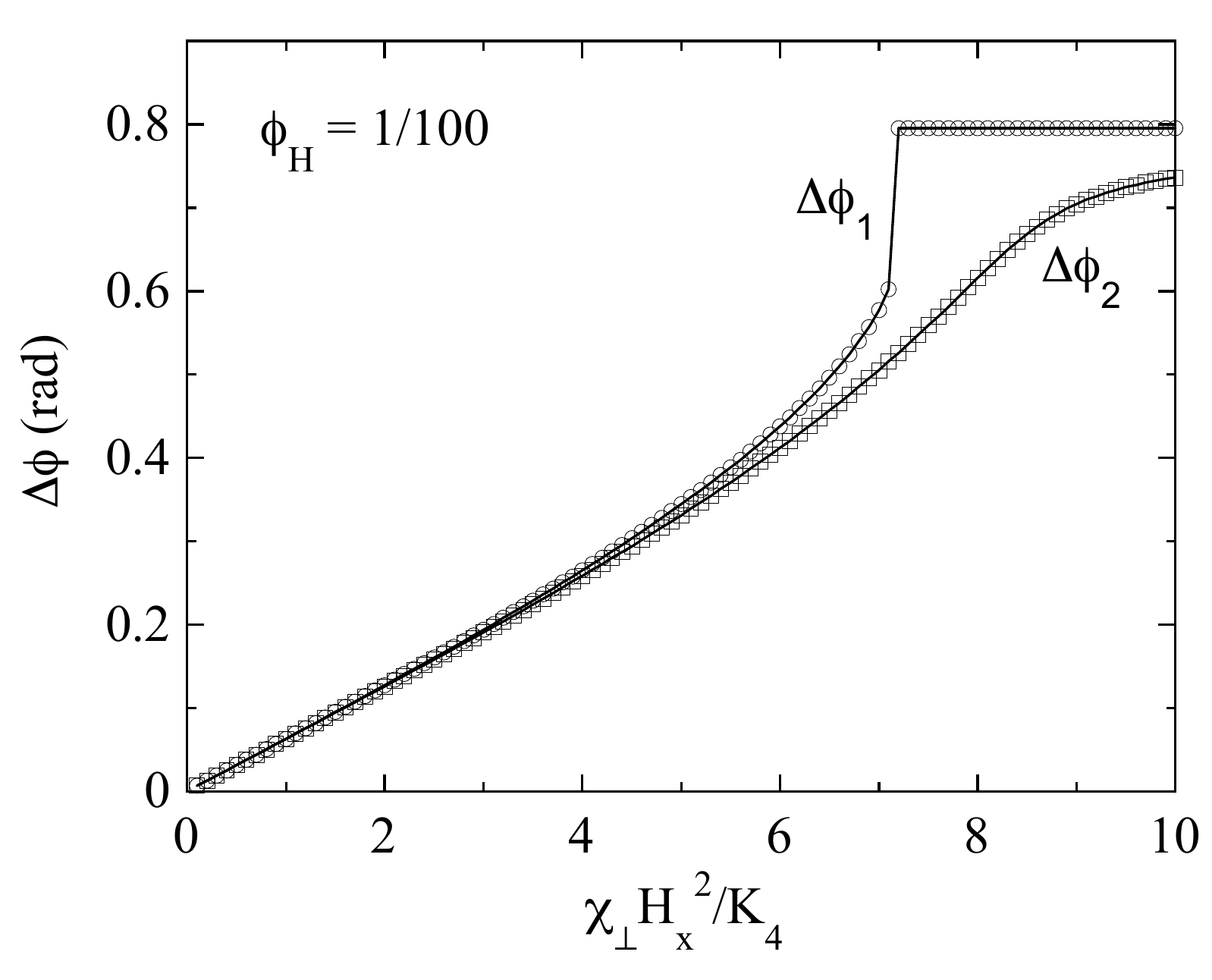}
\caption {The angles $\Delta\phi_1$ and $\Delta\phi_2$ in Eqs.~(\ref{Eqs:phiAB}) versus $\chi_\perp H_x^2/K$ for $\phi_H=1/100$~rad.  $\Delta\phi_1 = \pi/4$ for \mbox{$\chi_\perp H_x^2/K > 7.1$,} whereas $\Delta\phi_2$ eventually asymptotes to $\pi/4$~rad.}
\label{Fig_Delta_Phi_along_x}
\end{figure}

Thus, the total average energy at $T = 0$ normalized by $K_4$ is
\bea
E_{\rm ave}/K_4 &=& (E_{\rm anis\,ave} + E_{\rm mag\,ave})/K_4 \nonumber\\
&=& -\frac{1}{2}[\cos(4\phi_1) +  \cos(4\phi_2)\label{Eq:Etotal2}\\
&&-\frac{\chi_\perp H_x^2}{4K_4}\big\{2+\sin[2(\phi_1-\phi_H)]\nonumber\\
&&+\sin[2(\phi_2-\phi_H)]\big\}.\nonumber
\eea
Here we minimize of $E_{\rm ave}/K_4$ with respect to both $\Delta\phi_1$ and $\Delta\phi_2$ where $\phi_1$ and $\phi_2$ have maximum values of $\frac{\pi}{4}+\phi_H$ and $\frac{\pi}{4}-\phi_H$, respectively [see Fig.~\ref{Figure_domain}(a)].  As can be seen from the Fig.~\ref{Fig_Delta_Phi_along_x}, a discontinuous first-order transition is observed for $\Delta\phi_1$ for $\phi_H > 0$ at $H = H_{c1, \rm A}(\phi_H)$, where $\Delta\phi_1$ reaches $\frac{\pi}{4}+\phi_H$ in order for the moments to be perpendicular to $H_x$ at $H_{\rm c1,1}$. On the other hand, no such first-order transition is observed for $\Delta\phi_2$, where $\Delta\phi_2$ asymptotes continuously to $\frac{\pi}{4}-\phi_H$ at larger~$H$\@.   The critical field $H_{c1,1}$ is found to decrease and the discontinuity of $\Delta\phi_1$ at $H_{\rm c1,1}$ to increase  with increasing $\phi_H$\@.  These calculations reveal that when the applied field $H_x$ is not along a crystallographic $ab$-plane axis, the field responses of the moments in the two orthogonal domains A and B are quite different. Additional measurements along these lines would be of interest.

\section{\label{Sec:Conclu} Concluding Remarks}

The Eu square-lattice compound \eg\ exhibits \mbox{A-type} antiferromagnetic order at a N\'eel temperature $T_{\rm N} = 16.4(2)$~K with the moments aligned in the $ab$~plane.  A magnetic-field-induced anomaly is observed at low fields in the $M_{ab}(H)$ isotherms at $T=2~$K\@. We infer that the $H = 0$  A-type magnetic structure consists of four AFM domains associated with a fourfold in-plane magnetic anisotropy, where each domain consists of antiparallel moments in adjacent $ab$~planes along the $c$~axis.  On application of an in-plane magnetic field $H_x$, the collinear moments in each of the AFM domains gradually orient to become perpendicular to $H_x$ at a critical field $H_{c1}$, yielding a nonlinear $M_{ab}(H)$ at $T = 2$~K\@.  The $M_{ab}(H)$ behavior along the two $ab$-plane crystallographic symmetry directions [1,0,0] and [1,1,0] are quite different with respective critical fields $H_{c1, [1,0,0]} = 4.8$~kOe and $H_{c1, [110]} = 0.85$~kOe, respectively.  The experimental $M_{ab}(H)$ data for $H\parallel [1,0,0]$ and [1,1,0] were successfully modeled by a theory incorporating the fourfold tetragonal in-plane magnetic anisotropy and associated AFM domains.  However, the calculations predict a first-order transition when the in-plane field $H_x$ is at a finite angle to one of the [1,0,0] or [0,1,0] directions.  Since the theoretical calculations are done for $T = 0$~K whereas the experiments were performed at 2~K, it would be interesting to calculate the finite temperature effects on the $M(H)$ behavior.  In addition, the effects of the Eu-moment rotations for the small magnetic fields discussed here on the topological properties of \eg\ would be very interesting to explore.

Similar moment-reorientation effects due to small fields aligned in the $ab$~plane have recently been observed in the trigonal A-type antiferromagnets \emb\ and \ems\ containing $S = 7/2$ Eu$^{2+}$ spins with the moments aligned in the $ab$~plane~\cite{Pakhira2022b}.  It seems likely that $M(H)$ measurements for other layered Eu$^{2+}$ \mbox{spin-7/2} compounds with A-type AFM order and moments aligned in the layer plane would also exhibit low-field effects similar to those described here and in Ref.~\cite{Pakhira2022b}.

\acknowledgments

This research was supported by the U.S. Department of Energy, Office of Basic Energy Sciences, Division of Materials Sciences and Engineering.  Ames National Laboratory is operated for the U.S. Department of Energy by Iowa State University under Contract No.~DE-AC02-07CH11358.


\newpage

\begin{thebibliography}{99}

\bibitem{Jungwirth2018} T. Jungwirth, J. Sinova, A. Manchon, X. Marti, J. Wunderlich, and C. Felser, The multiple directions of antiferromagnetic
spintronics, Nat. Phys. {\bf 14}, 200 (2018).

\bibitem{Park2011} B. G. Park, J. Wunderlich, X. Mart\'{\i}, V. Hol\'{y}, Y. Kurosaki, M. Yamada, H. Yamamoto, A. Nishide, J. Hayakawa, H. Takahashi, A. B. Shick, and T. Jungwirth, A spin-valve-like magnetoresistance of an antiferromagnet-based tunnel junction, Nat. Mater. {\bf 10}, 347 (2011).

\bibitem{Fert2017} A. Fert, N. Reyren, and V. Cros, Magnetic Skyrmions: Advances in physics and potential applications, Nat. Rev. Mater. {\bf 2}, 17031 (2017).

\bibitem{Tokura2021} Y. Tokura and N. Kanazawa, Magnetic skyrmion materials, Chem. Rev. {\bf 121}, 2857 (2021).

\bibitem{Otrokov2019} M. M. Otrokov, I. I. Klimovskikh, H. Bentmann, D. Estyunin, A. Zeugner, Z. S. Aliev, S. Ga{\ss}, A. U. B. Wolter, A. V. Koroleva, A. M. Shikin, M. Blanco-Rey, M. Hoffmann, I. P. Rusinov, A. Yu. Vyazovskaya, S. V. Eremeev, Yu. M. Koroteev, V. M. Kuznetsov, F. Freyse, J. S\'{a}nchez-Barriga, I. R. Amiraslanov, M. B. Babanly, N. T. Mamedov, N. A. Abdullayev, V. N. Zverev, A. Alfonsov, V. Kataev, B. Büchner, E. F. Schwier, S. Kumar, A. Kimura, L. Petaccia, G. Di Santo, R. C. Vidal, S. Schatz, K. Ki{\ss}ner, M. \"{U}nzelmann, C. H. Min, Simon Moser, T. R. F. Peixoto, F. Reinert, A. Ernst, P. M. Echenique, A. Isaeva \& E. V. Chulkov, Prediction and observation of an antiferromagnetic topological insulator, Nature {\bf 576}, 416 (2019).

\bibitem{Rahn2018} M. C. Rahn, J. R. Soh, S. Francoual, L. S. I. Veiga, J. Strempfer, J. Mardegan, D. Y. Yan, Y. F. Guo, Y. G. Shi, and A. T. Boothroyd, Coupling of magnetic order and charge transport in the candidate Dirac semimetal EuCd$_2$As$_2$, Phys. Rev. B {\bf 97}, 214422 (2018).

\bibitem{Hirschberger2016} M. Hirschberger, S. Kushwaha, Z. Wang, Q. Gibson, S. Liang, C. A. Belvin, B. A. Bernevig, R. J. Cava, and N.~P. Ong, The chiral anomaly and thermopower of Weyl fermions in the half-Heusler GdPtBi, Nat. Mater. {\bf 15}, 1161 (2016).
    
\bibitem{Kurumaji2019} T. Kurumaji, T. Nakajima, M. Hirschberger, A. Kikkawa, Y. Yamasaki, H. Sagayama, H. Nakao, Y. Taguchi, T. Arima, Y. Tokura, Skyrmion lattice with a gianttopological Hall effect in a frustratedtriangular-lattice magnet, Science {\bf 365}, 914 (2019).
    
\bibitem{Jo2020} N. H. Jo, B. Kuthanazhi, Y. Wu, E. Timmons, T. H. Kim, L. Zhou, L. L. Wang, B. G. Ueland, A. Palasyuk, D. H. Ryan, R. J. McQueeney, K. Lee, B. Schrunk, A. A. Burkov, R. Prozorov, S. L. Bud'ko, A. Kaminski, and P. C. Canfield, Manipulating magnetism in the topological semimetal EuCd$_2$As$_2$, Phys. Rev. B {\bf 101}, 140402(R) (2020).

\bibitem{Riberolles2020} S. X. M. Riberolles, T. V. Trevisan, B. Kuthanazhi, T. W. Heitmann, F. Ye, D. C. Johnston, S. L. Bud’ko, D. H. Ryan, P. C. Canfield, A. Kreyssig, A. Vishwanath, R. J. McQueeney, L. L. Wang, P. P. Orth, and B. G. Ueland, Magnetic crystalline-symmetry-protected axion electrodynamics and field-tunable unpinned Dirac cones in EuIn$_2$As$_2$, Nat. Commun. {\bf 12}, 999 (2021).
    
\bibitem{Li2019} H. Li, S.-Y. Gao, S.-F. Duan, Y.-F. Xu, K.-J. Zhu, S.-J. Tian, J.-C. Gao, W.-H. Fan, Z.-C. Rao, J.-R. Hugang, J.-J. Li, D.-Y. Yan, Z.-T. Liu, W.-L. Liu, Y.-B. Huang, Y.-L. Li, Y. Liu, G.-B. Zhang, P. Zhang, T. Kondo, S. Shin, H.-C. Lei, Y.-G. Shi, W.-T. Zhang, H.-M. Weng, T. Qian, and H. Ding, Dirac Surface States in Intrinsic Magnetic Topological Insulators EuSn$_2$As$_2$ and MnBi$_{2n}$Te$_{3n+1}$, Phys. Rev. X {\bf 9}, 041039 (2019).

\bibitem{Shang2021} T. Shang, Y. Xu, D. J. Gawryluk, J. Z. Ma, T. Shiroka, M. Shi, and E. Pomjakushina, Anomalous Hall resistivity and possible topological Hall effect in the \ea\ antiferromagnet, Phys. Rev. B {\bf 103}, L020405 (2021).

\bibitem{Zhang2022} H. Zhang, X. Y. Zhu, Y. Xu, D. J. Gawryluk, W. Xie, S. L. Ju, M. Shi, T. Shiroka, Q. F. Zhan, E. Pomjakushina, and T. Shang, Giant magnetoresistance and topological Hall effect in the \eg\ antiferromagnet, J. Phys.: Condens. Matter {\bf 34}, 034005 (2022).

\bibitem{Moya2021} J. M. Moya, S. Lei, E. M. Clements, K. Allen, S. Chi, S. Sun, Q. Li, Y. Y. Peng, A. Husain, M. Mitrano, M. J. Krogstad, R. Osborn, P. Abbamonte, A. B. Puthirath, J. W. Lynn, and E. Morosan, Incommensurate magnetic orders and possible fieldinduced skyrmions in the square-net centrosymmetric EuGa$_2$Al$_2$ system, arXiv:2110.11935 (2021).

\bibitem{Kneidinger2014} F. Kneidinger, L. Salamakha, E. Bauer, I. Zeiringer, P. Rogl, C. Blaas-Schenner, D. Reith, and R. Podloucky, Superconductivity in noncentrosymmetric BaAl$_4$ derived structures, Phys. Rev. B {\bf 90}, 024504 (2014).

\bibitem{Araki2014} S. Araki, Y. Ikeda, T. C. Kobayashi, A. Nakamura, Y. Hiranaka, M. Hedo, T. Nakama, and Y. \={O}nuki, Charge density wave transition in EuAl$_4$, J. Phys. Soc. Jpn. {\bf 83}, 015001 (2014).

\bibitem{Nakamura2014} A. Nakamura, Y. Hiranaka, M. Hedo, T. Nakama, Y. Miura, H. Tsutsumi, A. Mori, K. Ishida, K. Mitamura, Y. Hirose, K. Sugiyama, F. Honda, T. Takeuchi, T. D. Matsuda, E. Yamamoto, Y. Haga, and Y. \={O}nuki, Unique Fermi surface and emergence of charge density wave in EuGa$_4$ and EuAl$_4$, Jpn. Phys. Soc. Conf. Proc. {\bf 3}, 011012 (2014).

\bibitem{Nakamura2015} A. Nakamura, T. Uejo, F. Honda, T. Takeuchi, H. Harima, E. Yamamoto, Y. Haga, K. Matsubayashi, Y. Uwatoko, M. Hedo, T. Nakama, and Y. \={O}nuki, Transport and magnetic properties of EuAl$_4$ and EuGa$_4$, J. Phys. Soc. Jpn. {\bf 84}, 124711 (2015).

\bibitem{Stavinoha2018} M. Stavinoha, J. A. Cooley, S. G. Minasian, T. M. McQueen, S. M. Kauzlarich, C.-L. Huang, and E. Morosan, Charge density wave behavior and order-disorder in the antiferromagnetic metallic series $\rm Eu(Ga_{1-x}Al_x)_4$, Phys. Rev. B {\bf 97}, 195146 (2018).

\bibitem{Ramakrishnan2022} S. Ramakrishnan, S. R. Kotla, T. Rekis, Jin-Ke Bao, C. Eisele, L. Noohinejad, M. Tolkiehn, C. Paulmann, B. Singh, R. Verma, B. Bag, R. Kulkarni, A. Thamizhavel, B. Singh, S. Ramakrishnan, S. van Smaalen, Orthorhombic charge density wave on the tetragonal lattice of EuAl$_4$, arXiv:2202.10282 (2022).

\bibitem{Nakamura2013} A. Nakamura, Y. Hiranaka, M. Hedo, T. Nakama, Y. Miura, H. Tsutsumi, A. Mori, K. Ishida, K. Mitamura, Y. Hirose, K. Sugiyama, F. Honda, R. Settai, T. Takeuchi, M. Hagiwara, T. D. Matsuda, E. Yamamoto, Y. Haga, K. Matsubayashi, Y. Uwatoko, H. Harima, and Y. \={O}nuki, Magnetic and Fermi surface properties of EuGa$_4$, J. Phys. Soc. Jpn. {\bf 82}, 104703 (2013).

\bibitem{Khanh2020} N. D. Khanh, T. Nakajima, X. Yu, S. Gao, K. Shibata, M. Hirschberger, Y. Yamasaki, H. Sagayama, H. Nakao, L. Peng, K. Nakajima, R. Takagi, T. Arima, Y. Tokura \& S. Seki, Nanometric square skyrmion lattice in a centrosymmetric tetragonal magnet, Nat. Nanotech. {\bf 15}, 444 (2020).

\bibitem{Hirschberger2019} M. Hirschberger, T. Nakajima, S. Gao, L. Peng, A. Kikkawa, T. Kurumaji, M. Kriener, Y. Yamasaki, H. Sagayama, H. Nakao, K. Ohishi, K. Kakurai, Y. Taguchi, X. Yu, T. Arima \& Y. Tokura, Skyrmion phase and competing magnetic orders on a breathing kagom\'{e} lattice, Nat. Commun. {\bf 10}, 5831 (2019).

\bibitem{Hayami2021} S. Hayami and Y. Motome, Square skyrmion crystal in centrosymmetric itinerant magnets, Phys. Rev. B {\bf 103}, 024439 (2021).

\bibitem{Zhu2022} X. Y. Zhu, H. Zhang, D. J. Gawryluk, Z. X. Zhen, B. C. Yu, S. L. Ju, W. Xie, D. M. Jiang, W. J. Cheng, Y. Xu, M. Shi, E. Pomjakushina, Q. F. Zhan, T. Shiroka, and T. Shang, Spin order and fluctuations in the \ea\ and \eg\ topological antiferromagnets: A $\mu$SR study, Phys. Rev. B {\bf 105}, 014423 (2022).

\bibitem{Lei2022} S. Lei, K. Allen, J. Huang, J. M. Moya, B. Casas, Y. Zhang, M. Hashimoto, D. Lu, J. Denlinger, L. Balicas, M. Yi, Y. Sun, and E. Morosan, Weyl nodal ring states and Landau quantization with very large magnetoresistance in square-net magnet \eg, 	arXiv:2208.06407 (2022).

\bibitem{Kawasaki2016} T. Kawasaki, K. Kaneko, A. Nakamura, N. Aso, M. Hedo, T. Nakama, T. Ohhara, R. Kiyanagi, K. Oikawa, I. Tamura, A. Nakao, K. Munakata, T. Hanashima, and Y. \={O}nuki, Magnetic structure of divalent europium Compound \eg\ studied by single-crystal time-of-flight neutron diffraction, J. Phys. Soc. Jpn. {\bf 85}, 114711 (2016).

\bibitem{Yogi2013} M. Yogi, S. Nakamura, N. Higa, H. Niki, Y. Hirose, Y. \={O}nuki, and H. Harima, $^{153}$Eu and $^{69,71}$Ga zero-field NMR study of antiferromagnetic state in \eg, J. Phys. Soc. Jpn. {\bf 82}, 103701 (2013).

\bibitem{Johnston2012} D. C. Johnston, Magnetic Susceptibility of Collinear and Noncollinear Heisenberg Antiferromagnets, Phys. Rev. Lett. {\bf 109}, 077201 (2012).

\bibitem{Johnston2015} D. C. Johnston, Unified molecular field theory for collinear and noncollinear Heisenberg antiferromagnets, Phys. Rev. B {\bf 91}, 064427 (2015).

\bibitem{Pakhira2020} S. Pakhira, M. A. Tanatar, and D. C. Johnston, Magnetic, thermal, and electronic-transport properties of ${\rm EuMg_2Bi_2}$ single crystals, Phys. Rev. B {\bf 101}, 214407 (2020).

\bibitem{Pakhira2021} S. Pakhira, T. Heitmann, S. X. M. Riberolles, B. G. Ueland, R. J. McQueeney, D. C. Johnston, and D. Vaknin, Zero-field magnetic ground state of ${\rm EuMg_2Bi_2}$, Phys. Rev. B {\bf 103}, 024408 (2021).

\bibitem{Pakhira2022b} S. Pakhira, Y. Lee, L. Ke, and D. C. Johnston, Magnetic-field-induced $ab$-plane rotation of the Eu magnetic moments in trigonal EuMg$_2$Bi$_2$ and EuMg$_2$Sb$_2$ single crystals below their N\'{e}el temperatures, arXiv:2208.06020 (2022).

\bibitem{Pakhira2021a} S. Pakhira, M. A. Tanatar, T. Heitmann, D. Vaknin, and D. C. Johnston, A-type antiferromagnetic order and magnetic phase diagram of the trigonal Eu spin-$\frac{7}{2}$ triangular-lattice compound EuSn$_2$As$_2$, Phys. Rev. B {\bf 104}, 174427 (2021).

\bibitem{Pakhira2022} S. Pakhira, F. Islam, E. \'{O}Leary, M. A. Tanatar, T.~Heitmann, Lin-Lin Wang, R.~Prozorov, A.~Kaminski, D.~Vaknin, and D.~C.~Johnston, A-type antiferromagnetic order in semiconducting \ems\ single crystals, Phys. Rev. B {\bf 106}, 024418 (2022).

\bibitem{Johnston2016} D. C. Johnston, Magnetic dipole interactions in crystals, Phys. Rev. B {\bf 93}, 014421 (2016).

\end{thebibliography}
\end{document}